# Cancer systems biology in the genome sequencing era: Part 2, evolutionary dynamics of tumor clonal networks and drug resistance

Edwin Wang[1,2,*], Jinfeng Zou[1,3,4] Naif Zaman[1,5], Lenore K. Beitel[3,4], Mark Trifiro[3,4] and Miltiadis Paliouras[3,4]

1. National Research Council Canada, Montreal, Canada
2. Center for Bioinformatics, McGill University, Montreal, Canada
3. Lady Davis Institute, Montreal, Canada
4. Department of Medicine, McGill University, Montreal, Canada
5. Department of Anatomy and Cell Biology, McGill University, Montreal, Canada

* Corresponding to EW (edwin.wang@cnrc-nrc.gc.ca)
6100 Royal Mount Ave, Montreal, QC, Canada, H4P 2R2
Tel: 1-514-496-0914
Fax: 1-514-496-5143

**Abstract**

A tumor often consists of multiple cell subpopulations (clones). Current chemo-treatments often target one clone of a tumor. Although the drug kills that clone, other clones overtake it and the tumor reoccurs. Genome sequencing and computational analysis allows to computational dissection of clones from tumors, while singe-cell genome sequencing including RNA-Seq allows to profiling of these clones. This opens a new window for treating a tumor as a system in which clones are evolving. Future cancer systems biology studies should consider a tumor as an evolving system with multiple clones. Therefore, topics discussed in Part 2 of this review include evolutionary dynamics of clonal networks, early-warning signals (e.g., genome duplication events) for formation of fast-growing clones, dissecting tumor heterogeneity, and modeling of clone-clone-stroma interactions for drug resistance. The ultimate goal of the future systems biology analysis is to obtain a 'whole-system' understanding of a tumor and therefore provides a more efficient and personalized management strategies for cancer patients.

**Keywords:**
Genome sequencing, systems biology, signaling network, cancer evolution, tumor heterogeneity, tumor clone, drug resistance, early-warning signal, personalized medicine

# 1. Introduction

The transformation from a normal cell into a tumor cell is a gradual evolution process in which genomic alterations accumulate in a step-wise manner. We described several models of tumorigenesis in Part 1 of this review [1]. These models suggest that for most of the tumors, tumorigenesis involves progression from early, slow-growing clones to late, fast-growing clones [1]. Although clones within a tumor are genetically related, they gain different growth or invasive capabilities so that they may have different response to a drug treatment.

In the past decade, cancer systems biology research has led to a series of discoveries and the development of new methods [2,3]. For example, network approaches have led to identification of high-quality cancer prognostic biomarkers [4-6] and drug target discovery and drug repositioning [7,8]; Network modeling of network modules and motifs has not only pinpointed biomarkers, but also provided insights into cancer therapies [9-18]. For example, analysis of signaling networks with cancer mutation [19,20] and cancer phosphoproteomic data [21] suggests that cancer signaling is highly enriched in the network regions which are defined by the hub kinases and hub kinase substrates. In addition, a series of methods for reverse-engineering of gene regulatory networks have been developed [22,23]. However, almost all of these studies have focused on different types of omic data derived from whole tumors. These data represent readouts of mixed clones from the tumors, and therefore, introduce lots of noise and make the network modeling inaccurate.

Advances in genome sequencing technology allows for computational dissection of clones and reconstruction of the evolutionary history of the tumors [1,24,25]. Emerging single-cell genome sequencing and RNA-Seq technologies allows to obtaining genomic alterations and gene expression profiles for individual clones. By accessing the omic data at the clone level, we could conduct systems biology studies of tumor clones. In Part 1 [1] of this review, we described the computational quantification of tumor subpopulations; clone-based network modeling, cancer hallmark-based networks and their high-order rewiring principles and the principles of cell survival networks of fast-growing clones. For example, network modeling (Zaman et al. unpublished observations) of cancer fast-growing clones uncovered the principles of the cancer cell survival signaling networks – a set of genes are recurrently used by genomic alterations (mutations and copy number variations) and cancer essential genes (i.e., knocking-down such a gene leads to cancer cell death). Theses cancer cell survival networks represent an end-point of cancer cell survival signaling machine. It would be interesting to know how genomic alterations drive cancer cells to converge to the cancer survival networks, what the differences are between the networks of the clones within a tumor, and how to treat a cancer patient and overcome drug resistance by managing the patient's tumor clones.

To answer these questions, it is necessary to model the evolutionary dynamics of clonal networks, clone-clone and clone-stroma interactions to understand how a tumor is evolving and how drug resistance emerges. Therefore, in Part 2 of this review we will discuss clonal network evolution, sharp transition warning signals from slow-growing to fast-growing clones, and how drug resistance could result from clonal backup and clone-stroma interactions. Understanding of these issues from a systems biology point of view will help in understanding of a tumor as 'a whole system' and finally in developing personalized strategies to manage cancer patients.

## 2. Evolutionary dynamics of clonal networks and early-warning signals of fast-growing clone formation

Tumorigenesis is typically viewed as a gradual evolution process, taking years to accumulate the multiple genomic alterations required to drive the cancer's aggressive growth. Genome sequencing of breast cancer and leukemia [24,26] suggests that mutational processes evolve across the lifespan of a tumor. As the cells accumulate many thousands of mutations, the developing cancer starts to diverge into clones of genetically related cells. By the time the cancer is diagnosed in a clinical setting, one of these clones has become the dominant population in the tumor, so that the tumor is clinically 'detectable' to doctors. These studies suggest that evolution holds the key to understanding why tumors often recur after treatment, and to the development of better therapies.

With new computational tools [27,28], we could dissect the mutations in the contexts of timing and clones. These data open a new opportunity to model the evolutionary dynamics of molecular networks of the clones within a tumor. The mutation and copy number variations (CNV) data could be used to construct clone-based cell survival networks. By modeling of the rewiring of these networks along the timing of clonal development, we could understand how and why these clones evolved and even predict new mutations based on a given clonal network. Previously, without using clonal information, a dynamic cascaded method (DCM), which is based on the intra-stage steady-rate assumption and the continuity assumption, has been used to reconstruct dynamic gene networks from sample-based transcriptional data for evolving networks [29]. Similar approaches could be applied to tumor clones. In addition, by taking into account the default genetic profiles of cell of origins, germline variants (e.g., derived from GWAS, genome-wide association studies), and system-constraints (see Part 1 of this review [1]), it is possible to infer specific combinatory patterns of genomic alterations in networks. By doing so, we could identify recurrent network modules which represent preferred clonal evolutionary paths. Different clonal evolutionary paths can be translated into predictions for predisposition and drug intervention.

The process of cancer initiation and progression is a natural experimental evolutionary system. The evolutionary process of cancer cells is highly dynamic. In general, a wide range of complex systems including physics, physiology, ecology and social sciences have critical transitions. It is becoming increasingly clear that many complex systems have critical thresholds, so-called tipping points, at which the system shifts abruptly from one state to another [30]. For cancer cells, the surprising shift that occurs during the cancer cell evolution is marked by the sharply different states between a fast-growing clone and its direct mother clone. Evolutionary studies via genome sequencing suggest that fast-growing clones come late and are derived from slow-growing clones of the early stage in tumorigenesis. Previous studies suggest that mutations in cancer cell evolution play an additive/accumulative role in a small-scale and gradual manner [31]. However, based on evolutionary studies via genome sequencing, we expect that certain genomic alterations drive a sharp transition between a fast-growing clone and its direct mother (a slow-growing clone). In this regard, it is interesting to model and compare clonal networks, especially between fast-growing clone and its direct mother. Such a study could reveal early-warning signals for forming fast-growing clones. In the past, most network evolution studies have focused on single networks or comparisons of networks of different species [21,32-34]. For tumor clonal evolution, we could focus on time-course networks (i.e., networks reflect the time series of clonal evolution within a tumor).

Without the expansion of a fast-growing clone, a tumor can't be formed. If we could detect the early-warning signals for the sharp transition, cancer prevention strategies could be applied at this stage [35,36]. Theoretically, highly complex systems such as ecosystems have shown expected early-warning signals [30]. Sharp transitions are related to 'catastrophic bifurcations', where, once a threshold is exceeded, a positive feedback pushes the system through a phase of directional change towards a contrasting state [30]. Capturing the essence of shifts at tipping points in cell signaling pathways has been attempted [37]. To model cancer clonal evolution and identify potential early-warning signals, the networks should reflect the relations of genomic alterations and cell proliferation functions – cell proliferation, cell cycle, and apoptosis. Some positive regulatory loops or positive network feedback motifs could encode the early-warning signals. Positive feedback is widely observed in complex systems, ranging from cellular circuits to ecosystems. A handful of evidence has shown that positive feedback leads to alternative stable states and tipping points in various ecological systems. Furthermore, such loops might be organized into a set of bi-stable or multi-stable circuits exhibiting switch-like behavior. Bistable switch networks could be constructed using pairs of genes with double-negative feedback. The ON (upregulated)/OFF (downregulated) states could be used to model the transition [38-40]. It may be interesting to examine the recurrent positive feedback network motifs or functional modules during the sharp transitions. During evolution, gradually rewiring (i.e., adding new genomic aberrations and then recruiting new genes) of the clonal networks could gradually increase the power of positive feedback network motifs/modules until the threshold is reached, such that an extra event of genomic alteration will push a sharp transition to form a fast-growing clone. Early-warning signals (i.e., recurrent positive feedback network motifs/modules) could be dependent on the cell default state (e.g., cell of origin and germline variants), early genomic alteration events such as mutation of P53, Ras or EGFR, and the final key genomic alteration event which generates a fast-growing clone from its direct mother clone. In addition, early-warning signals could be represented by mutational signatures, which are indicative of genomic mutation patterns, or gene expression signatures, which represent gene expression changes during the transition. Some efforts have been made to look for early-warning signals of diseases, but experimental validations for these signals are still lacking [41].

In fact, we suspect that genome duplication event is most likely to be an early-warning signal (at least for breast cancer): (1) by analysis of 16 breast cancer cell lines, we found that cancer essential genes (i.e., knocking down such a gene will lead to cancer cell die or very slow-growing) have been enriched more than 50 times on average in amplified genes than driver-mutating genes; (2) from the breast cancer genome sequencing data generated by Nik-Zainal et al. [24] (Fig 1), we noted that most (10/15) of the tumor genomes experienced at least one genome duplication event which often leads to massive gene amplifications and deletions. Importantly, this event is often the last round of the gene amplification events and the late stage (fast-growing clones occur in late stage too) during tumor development. These data support our hypothesis that the accumulation of a certain number of amplified genes is critical for driving tumorigenesis. Based on these results, we suspected that genome duplication could be the rate-limiting step for tumor development, and therefore could be an early-warning signal for fast-growing clone formation. Network modules representing early-warning signals could be identified by differential network analysis of the networks which represent the clone before and after genome duplications. Furthermore, understanding of factors such as certain combinations of mutations or other environmental or epigenetic factors, which trigger genome duplication, could shed light on cancer prevention.

**3. Dissecting heterogeneity and modeling of drug resistance for personalized treatment**

Targeted cancer therapy is promising, however, in general only 20-30% of cancer patients respond to drug treatments in today's clinical practice. The emergence of drug resistance in the course of treatment remains a major challenge in cancer therapy. Heterogeneity has been proposed as one of the major reasons for the failure of drug treatment in cancer management. Tumor genome sequencing suggests that a high degree of cancer cell heterogeneity exists in each tumor. Thus, one common source of drug resistance comes from the presence of multiple clones. Clinically, it is well-known that despite several different treatments, each somewhat successful at first, that tumors grow back again. It has been suggested that the drug kills one fast-growing clone (usually the dominant clone), but other fast-growing clones overtake it and the tumor reoccurs. By sequencing chronic lymphocytic leukemia (CLL) tumor genomes before and after chemotherapy, researchers found that patients whose original leukemia harbored clones with one or more cancer-driver genes often died sooner than patients without multiple clones [26]. Some fast-growing, but non-dominant, clones may have a fairly minimal presence before treatment and predominate after treatment. The clones that originally were somewhat rare or non-dominant may have gained a competitive advantage for proliferation and growth.

To overcome heterogeneity-derived drug resistance, it is critical to dissect the clones and model their networks (clone-based networks) for cell survival (cancer hallmark-based networks). By modeling the cell survival networks of the fast-growing clones of a tumor, we could identify key genes as drug targets. It is unclear whether common drug targets exist for multiple fast-growing clones within a tumor. It is possible that the late-occurring fast-growing clones gain extra genomic alterations which could backup (i.e., redundant functional pathways) the targets of its parental clones. It is necessary to model the backup within a clonal network, where a network component could functionally replace another one, especially in terms of cell survival and proliferation. For example, different gene alterations within the same pathway and cooperation of pathways perturbed by mutations can lead to the same phenotype. Finding the co-altered functional modules by integrating of mutations, CNVs and gene expressions could model network backup [42]. For generating a same phenotype, if Pathway A cooperates with Pathway B, and Pathway A cooperates with Pathway C, then Pathways B and C could be functionally backup each other. The synthetic lethality concept has been also explored for modeling of the functional redundancy within a network. For example, the synergistic outcome determination (SOD) approach, which constructs a synergistic network based on gene expression data and cancer prognostic information, has been used for performing module analysis to discriminate drugs from a broad set of test compounds and revealing the mechanisms of drug combinations [43]. The combinatorial perturbation approach, which constructs network models from perturbed molecular profiles assuming that after perturbation the system evolves according to nonlinear differential equations, has been used for identifying drug pairs to overcome network backup [44]. The current backup modeling approaches are still in their infancy, it is necessary to develop more advanced methods which could predict backup more accurately and more comprehensively.

Current cancer treatments do not take clonal diversity into account and often target only the dominant fast-growing clone. Such an approach leaves the possibility that one of the minor fast-growing clones will then replicate and become dominant, leading to re-occurrence of the tumor. Thus, modeling of the networks of both minor and dominant fast-growing clones within a tumor could provide a pivotal role in treating destructive cancers in the most efficient way. Many network methods for finding cancer genes or drug targets have been developed for a single network. One approach is modeling of networks by defining seed genes. These methods include predicting drug targets using metabolic networks [45], ranking genes based on PageRank concept (e.g., NetRank, [46], defining centrality measures according to their relevance to the seed genes in the network (e.g., NetworkPrioritizer, [47]), employing random walks (e.g., NetWalker, a context-specific random walker [48]), or using a RVM-based ensemble model (TARGETgene, [49]). Another approach is performing an integrative analysis of mutations and CNVs on networks [50,51] or constructing causal-target networks using gene expression and CNV/mutations (e.g., using differentially expressed genes with CNVs to determine paths from causal alterations to these target genes based on network topology [52] or checking the mutations on the interaction interfaces between protein interactions [53]. Network perturbation has been explored to identify drug targets. For example, in silico perturbation of the receptors of the networks [54] or Boolean network perturbing of networks [55] have been used for finding drug targets. Karlebach and Shamir [56] used a network perturbation method to find the smallest perturbations on a network formulated as a Petri net which can yield a desired phenotype.

Although many methods have been explored, there is still room for improving the accuracy of the predictions. By modeling clonal networks, it is possible to predict drug targets for each clone. If clones within a tumor do not share common targets, it would be advantageous to identify multiple drug targets for individual. In this situation, combinatory therapy should be applied. Such an approach, which takes all the fast-growing clones of a tumor into account, could help us tailor our therapy to those specific clones, and better predict which patients are likely to relapse. Moreover, it could help in developing novel therapeutic/patient-management paradigms that address the cancer evolutionary landscape and clonal diversity.

Another common source of drug resistance comes from tumor microenvironments or stroma. Tumors are surrounded by multiple supportive cell types. Anticancer drugs that are capable of killing tumor cells are frequently rendered ineffective when the tumor cells are cultured in the presence of stromal cells. Straussman and colleagues [57] used a co-culture system in which 45 cancer cell lines were cultured alone or with 1 of 23 stromal cell lines in the presence of 35 oncology drugs. They discovered that HGF, a ligand for the receptor tyrosine kinase (RTK) MET caused the resistance to a BRAF inhibitor (PLX4720). Validation experiments with HGF-neutralizing antibodies showed that HGF was both necessary and sufficient to confer drug resistance. They found stroma-mediated resistance was common in targeted agents. Overall, there was evidence of microenvironment-mediated resistance in up to 65% of the targeted agents studied. Similarly, Wilson and colleagues [58] showed that HGF attenuated the response of MET-expressing melanoma cells to a BRAF inhibitor, and inhibition of MET blocked HGF-induced resistance *in vitro* and *in vivo*. These findings suggest that stroma is an important source of anticancer drug resistance. Modeling of stromal-mediated resistance may provide a hitherto untapped strategy for overcoming drug resistance.

Tumor-stromal communications mainly rely on signaling transduction mechanisms via ligand-receptor interactions, i.e., ligands secreted by stroma can activate receptor-dependent pathways of tumor cells. Specific ligands secreted by stroma can promote drug resistance to a given drug. If we are able to predict the specific ligand-receptor interaction(s) that are likely to promote stroma-mediated resistance for a given drug, then we can predict which novel combinatorial therapy can be used for a tumor, i.e., which antibody will likely block stroma-mediated resistance and therefore sensitize the tumor to a specific drug. Clearly, modeling of the interactions between the signaling networks of fast-growing clones within a tumor and the stromal-signaling network could provide hints about resistance to the drugs that are used for treating each clone.

As multiple clones co-exist in a tumor (Fig 2), they undoubtedly have relationships in terms of genetic profiles: (1) one clone could support the growth of other clones, for example, a clone could amplify a ligand such as FGF, which could trigger FGF signaling pathways in other clones;  or a clone could interact with the tumor microenvironment to protect itself and other clones within the tumor from host immune systems; (2) one clone could suppress another clone's growth by either secreting inhibiting factors or  by using a larger portion of the available nutrients and growing aggressively to take over a large volume/space within a tumor; and (3) the clones grow independently and have no interactions with each other. Therefore, in addition to modeling clone-stroma network interactions, we also need to model clone-clone-stroma network interactions.

In summary, three levels of systems backups (i.e., functional redundancy) confer drug resistance: (1) new genomic alterations in late-occurring fast-growing clones could provide backup in network level so that a drug target in its parental clones could be not a target for the late-occurring fast-growing clones anymore; (2) the diversity of the fast-growing clones within a tumor provides backup in the manner of ecological population dynamics; (3) interactions of clone-clone-stroma could provide backup at the host level. Better understanding of the backup at these levels should help develop new insights into how to tackle the problem of cancer. Ultimately, this will lead to new, more personalized treatments that will improve patient care. For example, tumor samples could be used for sequencing and "omic"-profiling, then data could be modeled using a systems approach, and finally combinatory drug targets will be proposed. The same tumor samples can be used to generate corresponding patient-derived mouse models for drug testing. By doing so, we could generate a 'whole-system' understanding of a tumor and provide a more efficient and personalized patient management strategy.

**4. Integrative network modeling**

One of the advantages of the systems approach is that multiple types of data could be integrated into one network and thus, integrative network modeling conducted. Although cancer has been recognized as a mutating disease, we cannot only focus on gene mutations. Almost all tumor genome sequencing papers have mainly discussed gene mutations. For example, they discussed which genes were highly mutated in samples, and even inferred major signaling pathways based purely on gene mutation information in the pathways [59-61]. These works often ignored many other factors such as CNVs, non-coding RNAs and so on. For instance, more than 40% of the genes in each tumor genome have been either amplified or deleted, whereas less than 10% of the genes are functionally mutated. Given the fact that many more genes are affected by CNVs than by mutations, amplified or deleted genes could play more important roles than mutated genes in clonal evolution and tumorigenesis. Similarly, the number of alterations in non-coding regions is proportionately higher than the number affecting coding regions. So are the numerous epigenetic changes in cancers. Integrative network modeling has been applied in cancer studies, for example, in constructing miRNA and post-translational networks [62-65], CNV-methylation-miRNA networks [66,67] or networks containing genes which are not only modulated but also mutated [68]. It is a worthy goal to transfer these analyses into clone-based networks and also consider emerging data types such as GWAS and single cell genome sequencing data. There are a growing number of massive international scientific collaborations such as Collaborative Oncological Gene-environment Study (COGS) [69,70] for conducting GWAS studies. In addition, new single cell genome sequencing technologies are being developed. Single cell genome sequencing could help in generating high-quality data for clones, and even be applied to circulating tumor cells. By integrating all these diverse data, we could model cancer tumors more comprehensively and finally develop effectively management strategies for cancer patients.

**Acknowledgements**
This work is supported by Genome Canada and Canadian Institutes of Health Research.

**Figure Legends**

**Figure 1. Evolutionary timing of gene duplication events during tumorigenesis**. The figure is modified from Nik-Zainal at al. [24]. Each line represents the timing of gene duplication events for each tumor. Most of the tumors have experienced genome duplication events, however, each tumor has experienced only one round of genome duplication, which is also the last gene duplication event.

**Figure 2. Interactions between clones and stroma**. A tumor often contains several fast-growing clones which could have interactions. The clones could also interact with stroma as well.

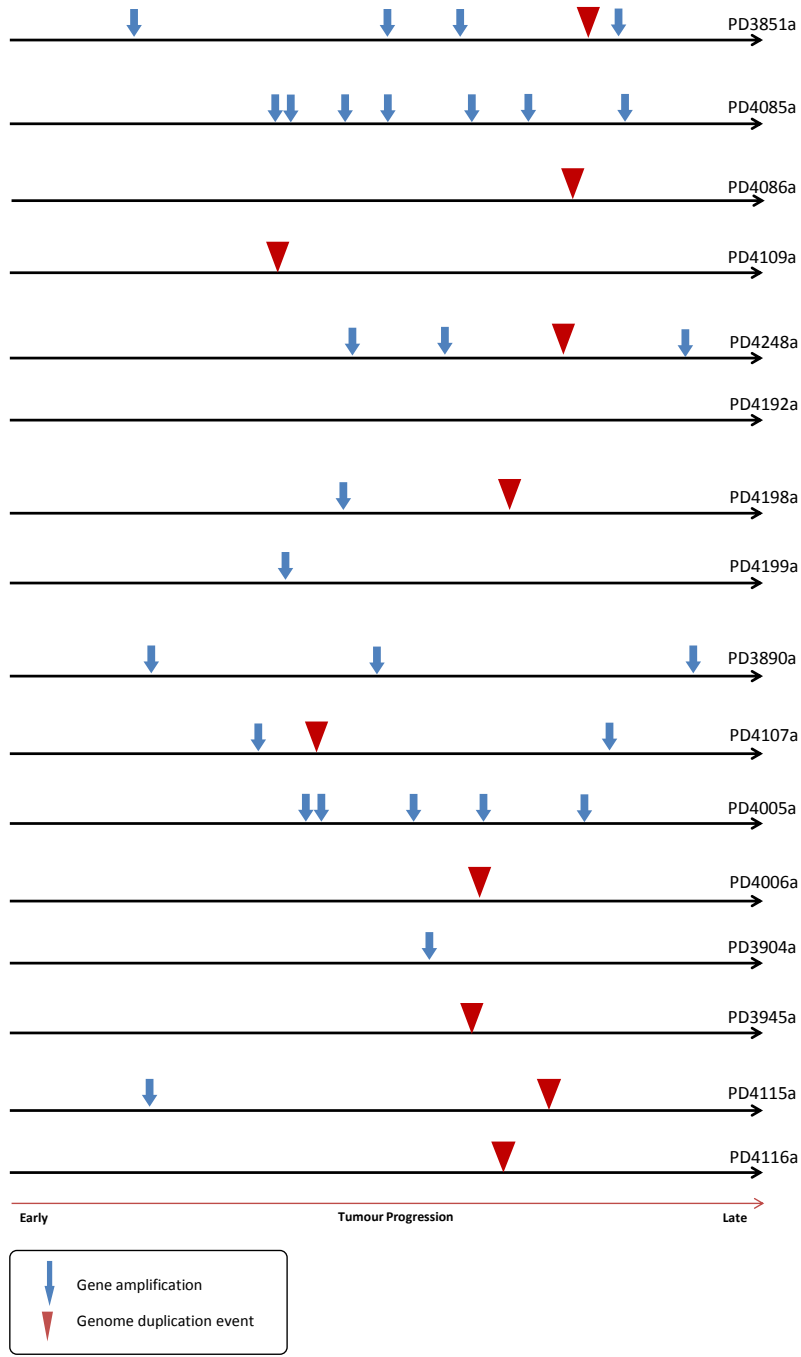

Fig 1

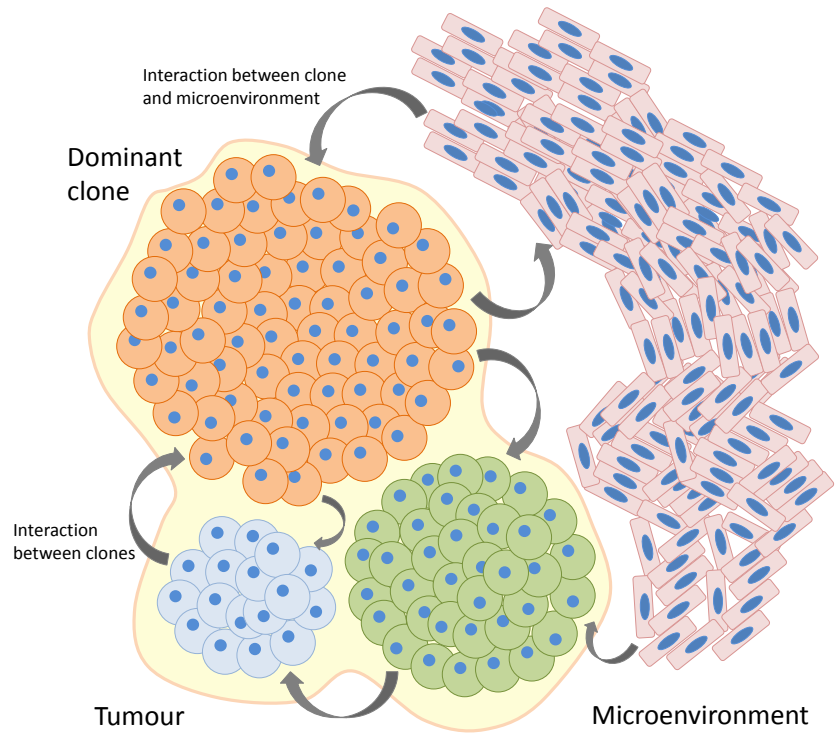

Fig 2